\documentclass[aps,prd,11pt,floatfix,groupedaddress]{revtex4}
\usepackage[dvips,final]{graphicx}
\usepackage{amsmath}
\usepackage{amssymb}
\usepackage{url}
\usepackage{subfigure}
\usepackage{textcomp}

\begin{document}

\title{Edge Detection, Cosmic Strings and the South Pole Telescope}

\author{Andrew Stewart}
\email[]{stewarta@physics.mcgill.ca}
\author{Robert Brandenberger}
\email[]{rhb@physics.mcgill.ca}

\affiliation{Physics Department, McGill University, Montreal, QC, H3A 2T8,
CANADA}

\date{\today}

\begin{abstract}
We develop a method of constraining the cosmic string tension $G\mu$ which 
uses the Canny edge detection algorithm as a means of  searching CMB 
temperature maps for the signature of the Kaiser-Stebbins effect. We test the 
potential of this method using high resolution, simulated CMB temperature 
maps. By modeling the future output from the South Pole Telescope project
(including anticipated instrumental noise), we find that cosmic strings
with $G\mu > 5.5\times10^{-8}$ could be detected.
\end{abstract}

\maketitle

\section{Introduction}

At very early times it is believed that the universe underwent a series of 
symmetry breaking phase transitions which led to the formation of different 
types of topological defects. Among them are linear topological defects known 
as cosmic strings (for reviews see e.g. 
\cite{2000csot.bookV,Hindmarsh:1994re,Brandenberger:1993by}). After creation, 
the cosmic strings form a random network of infinite strings and closed 
string loops, the arrangement of which evolves over time through string 
interactions. Cosmic strings can also have self-interactions that lead to 
the formation of closed loops via the exchanging of endpoints, or 
\emph{intercommutation} \cite{Shellard:1987bv}. When formed, cosmic string 
loops break off of the longer segments and continue to oscillate, losing 
energy via gravitational radiation, until eventually decaying. Infinitely 
long strings, on the other hand, cannot decay into gravitational radiation 
and survive indefinitely. The string network eventually approaches a scaling 
regime in which the number of strings crossing a given Hubble volume is fixed 
and the strings contribute some fraction of the total energy in the universe. 
The existence of a scaling solution is supported by independent numerical 
simulations of the evolution of the cosmic string network 
\cite{Albrecht:1984xv,PhysRevLett.60.257,Allen:1990tv,Albrecht:1989mk}. 
The quantity 
which characterizes the cosmic strings is their tension, $\mu$, which is 
equivalent to the mass per unit length. This tension is directly determined 
by the energy scale of the symmetry breaking during which the cosmic strings 
were formed. It is possible that cosmic strings could have been formed at 
many different epochs, meaning the tension of the cosmic strings can take a 
wide variety of values. When discussing cosmic strings it is more common to 
work with the dimensionless parameter $G\mu$, where $G$ is Newton's constant. 
Until the late 1990's, cosmic strings were studied as potential seeds for 
structure formation \cite{Turok:1985tt,Sato,Stebbins}. The eventual 
discovery of the acoustic peaks \cite{Boomerang,WMAP} in the angular 
power spectrum of the CMB lead to cosmic strings being ruled out as the 
main origin of structure in favour of the inflationary paradigm, since the 
angular power spectrum predicted by cosmic strings consists of only a single 
broad peak \cite{Periv,Albrecht,Turok}. Despite this, there currently 
exists a renewed interest in cosmic strings fueled by the study of different 
cosmological models in which their formation is generically predicted 
(see \cite{Brandenberger:1988aj,Jones:2003da,Jeannerot:2003qv} for just a 
few possibilities). It has also recently been shown that a contribution of 
less than 10\% of the observed CMB power on large scales coming from 
cosmic strings is acceptable \cite{Pogosian:2003mz,Wyman:2005tu,Fraisse:2006xc,Seljak:2006bg,Bevis:2006mj,Bevis:2007gh,Pogosian:2008am}. 

The current bounds on the cosmic string tension come from a variety of 
measurements. The gravitational waves emanating from many string loops at 
different times produce a stochastic background which is the focus of 
current interferometer and pulsar timing experiments. Pulsar timing, 
specifically, places a bound $G\mu<10^{-7}-10^{-8}$ on the cosmic string 
tension \cite{Damour:2004kw,Jenet:2006sv}. However, we note that in order 
to place a bound on $G\mu$ using gravitational wave constraints one must 
make assumptions about the size of the loops which are formed in the string 
network, the probability that strings will intercommute when crossing, and 
even the string model under consideration. Whereas the scaling solution
for the long string network is well established, the distribution of
loops is uncertain by several orders of magnitude. Therefore, the strength 
of the bounds obtained by considering gravitational radiation from
string loops can be questioned. A more robust bound on the tension comes 
from the angular power spectrum of the CMB (since the spectrum obtains
an important contribution from the long string network). Assuming
a scaling solution of the long string network with the parameters from
numerical studies of cosmic string evolution, the string contribution to 
the angular power spectrum of the CMB was determined, and the results of
these studies translate directly into a bound $G\mu<5\times10^{-7}$ 
\cite{Wyman:2005tu,Fraisse:2005hu}.

\begin{figure}
\begin{center}
\includegraphics[width=0.75\linewidth]{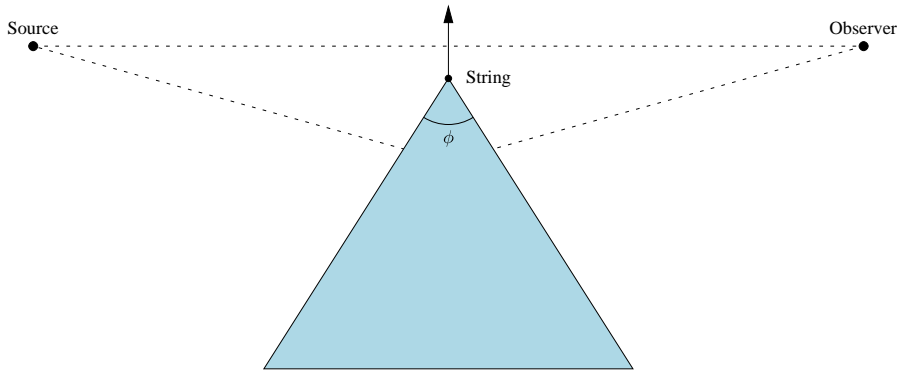}
\caption{\label{string} The geometry of the space-time near a cosmic string. 
Shown here is a slice of the space-time perpendicular to the orientation of 
the string. The coloured area represents a missing wedge with deficit angle 
$\phi$, while the dashed lines represent the paths of photons travelling from 
a source to an observer and the arrow shows the direction of motion of the 
string.}
\end{center}
\end{figure}

Along with the above mentioned phenomena, there exists another observational 
signature unique to cosmic strings which could be directly detected, namely, 
linear discontinuities in the temperature of the CMB. This signature was 
first studied by Kaiser and Stebbins \cite{Kaiser:1984iv} and is usually 
referred to as the KS-effect. This effect occurs because the space-time around 
a straight cosmic string is flat, but with a wedge, whose vertex lies along 
the length of the string, removed. The angle subtended by the missing wedge, 
$\phi$, is determined by the tension of the cosmic string as 
\cite{Vilenkin:1981zs}
\begin{equation}
\phi=8\pi G\mu\,.
\end{equation}
For an observer looking at a source while a cosmic string is moving 
transversely through the line of sight between the two, the photons passing 
from the source to the observer along one side of the string will appear to 
be Doppler shifted relative to those passing along the other side due to this 
non-trivial geometry (see Figure \ref{string}). If the source that the 
observer is viewing happens to be the CMB, this effect will manifest itself 
as discontinuities in the microwave background temperature along curves in 
the sky where strings are located. The magnitude of the step in temperature 
across a cosmic string is
\begin{equation}\label{KS}
\frac{\delta T}{T}=8\pi G\mu\gamma_sv_s\,|\hat{k}\cdot(\hat{v_s}\times\hat{e_s})|\,,
\end{equation}
where $v_s$ is the speed with which the cosmic string is moving, $\gamma_s$ 
is the relativistic gamma factor corresponding to the speed $v_s$, $\hat{v}_s$ 
is the direction of the string movement, $\hat{e}_s$ is the orientation of the 
string and $\hat{k}$ is the direction of observation \cite{Moessner:1993za}.
For cosmic strings formed in a phase transition in the early universe, the
``missing wedge'' produced by a string has, at time $t$, a finite depth
given by the Hubble radius $H^{-1}(t)$ at that time \cite{Joao}.
Some work has already been dedicated to searching for the KS-effect in current 
CMB data \cite{Jeong:2006pi,Lo:2005xt}, but a cosmic string signal was not 
found, leading to a constraint on the tension $G\mu\lesssim4\times10^{-6}$.

In this work we implement a method of detecting the temperature 
discontinuities in the CMB produced by cosmic strings via the KS-effect 
using an edge detection algorithm commonly employed in image analysis,
the Canny algorithm \cite{Canny:1986aa}. The motivation behind this choice is clear since the cosmic strings literally appear as edges in the CMB temperature. Depending on the sensitivity of the 
edge detection algorithm to the temperature edges, a bound on the cosmic 
string tension could then be imposed. This work is a continuation of a 
previous study \cite{Amsel:2007ki} which indicated that an edge detection 
method may lead to a significant improvement in the sensitivity to the
presence of cosmic strings compared to previous direct searches
for strings in the CMB. In this paper we improve the method proposed in 
\cite{Amsel:2007ki} and we investigate its application to surveys with a 
different set of specifications than those examined in that initial work.

We are interested in the cosmic strings in the network that survive until 
later times. The times relevant to the production of an edge signature in 
the CMB are the time of last scattering until the present day. Based on the 
evolution of the network, cosmic strings are more numerous around the time of 
last scattering than later times. On today's sky, those strings correspond to 
an angular scale of approximately 1$^\circ$. Therefore, an observation of the 
CMB with an angular resolution significantly less than 1$^\circ$ is necessary 
in order to be able to detect the edges related to these strings. With this 
in mind, we also focus on the application of the edge detection method to 
high resolution surveys of the CMB, particularly the future data from the 
South Pole Telescope.

The South Pole Telescope (SPT) \cite{Ruhl:2004kv} is a 10m diameter telescope 
being deployed at the South Pole research station. The telescope is designed 
to perform large area, high resolution surveys of the CMB to map the 
anisotropies. The SPT is designed to provide 1$^\prime$ resolution in the 
maps of the CMB, making it ideal to search for the KS-effect. Based on 
previous results \cite{Amsel:2007ki}, we believe that with such high 
resolution data our method could provide bounds on the cosmic string tension 
competitive with those of pulsar timing.

The remainder of this paper is arranged as follows: In section \ref{secmap}, 
we discuss the CMB maps used in our analysis with a focus on the anisotropies 
coming from Gaussian fluctuations and cosmic strings. In Section 
\ref{seccanny}, we outline the edge detection algorithm we are using, 
highlighting the details of our particular implementation. In Section 
\ref{seccount}, we discuss how we quantify the edge maps output by the 
edge detection algorithm and we explain the statistical analysis used to 
determine if a significant difference has been detected. In Section 
\ref{secresults}, we present the results of running the edge detection 
algorithm on CMB maps and the possible constraints on the cosmic string 
tension that could be applied. We finish in Section \ref{secdiscuss} with 
a discussion of our results.

\section{Map Making}
\label{secmap} 

For this initial investigation of edge detection as a method for constraining 
or even detecting cosmic strings, we generate CMB temperature anisotropy maps 
by means of numerical simulations and use these as the input for the edge 
detection algorithm. The simulated maps are constructed through the 
superposition of different temperature anisotropy components based on the type 
of effects being reproduced. We are interested in the simulation of small 
angular scale patches of the microwave sky, so we employ the flat-sky 
approximation \cite{White:1997wq}. In this approximation, the geometry of a 
small patch on the sky can be considered to be essentially flat. Thus, each 
map component, as well as the final map itself, is a two dimensional square 
image characterized by an angular size and an angular resolution. 
Specifically, we work with a square grid that has a size corresponding to the 
angular size being simulated, and a pixel size corresponding to the angular 
resolution being simulated. The pixels in the grid are indexed by two 
dimensional Cartesian coordinates $(x,y)$ and we take the upper left corner 
of the grid to be the origin.

The common component in every simulated CMB map is a set of temperature 
anisotropies produced by Gaussian inflationary fluctuations. We simulate 
these Gaussian fluctuations such that they account for all of the observed 
power in the CMB. Thus, in the absence of any other sources the final 
simulated map is simply equivalent to the Gaussian component and is consistent 
with observations. That is, we define
\begin{equation}
T(x,y) \, \equiv \, T_G(x,y)\,,
\end{equation}
where $T(x,y)$ represents the the final temperature anisotropy map and 
$T_G(x,y)$ represents the Gaussian component.

To make a CMB map including the effects of cosmic strings, we simulate a separate component of string induced temperature fluctuations produced via the KS-effect. In linear perturbation theory, if there are two sources of fluctuations, the resulting temperature anisotropies are given by a linear superposition of the individual sources. Therefore, the total temperature map is obtained by simply summing the contributions from Gaussian noise and from cosmic strings. Note, however, that the power of the Gaussian component must be adjusted in order to obtain the total observed power of CMB fluctuations. That is, if $T_G(x,y)$ is the WMAP-normalized Gaussian signal, then, in a map including cosmic strings, it needs to be reduced by a scaling factor $\alpha$, the value of which is determined by the tension of the cosmic strings being simulated. Denoting the string component by $T_S(x, y)$, the total temperature map is
\begin{equation}
T(x,y) \, \equiv \, \alpha\, T_G(x,y) + T_S(x,y)\,.
\end{equation} 
%
In this way the strings can contribute a fraction of the total power, while the final map is still in agreement with current measurements of the angular power spectrum of CMB anisotropies.  

Let us comment in more detail on the nature of this scaling. We demand that 
the angular power of the final combined temperature map match the observed 
angular power for multipole values up to the first acoustic peak, i.e. 
$l\lesssim220$. We choose this multipole range because it is tightly 
constrained by current observations \cite{Komatsu:2008hk}. However, as 
mentioned above, the Gaussian component alone accounts for all of the observed 
angular power in the CMB. Thus, this demand is equivalent to requiring that 
the angular power of the combined map match that of a pure Gaussian component. 
Working in the flat-sky approximation allows us to replace the usual spherical 
harmonic analysis of the CMB fluctuations by a Fourier analysis 
\cite{White:1997wq}. We can then express our condition as
\begin{equation}\label{power}
\langle|T_G(k<k_p)|^2\rangle \, = \, \alpha^2\langle|T_G(k<k_p)|^2\rangle 
+ \langle|T_S(k<k_p)|^2\rangle\,,
\end{equation}
where $k_p$ is the wavenumber corresponding to the first acoustic peak of the 
angular power spectrum of the CMB, $\langle|T_S(k<k_p)|^2\rangle$ is the 
average of the Fourier temperature anisotropy values from the string component 
for wavenumbers less than $k_p$ and $\langle|T_G(k<k_p)|^2\rangle$ is the 
equivalent object for the Gaussian component. From Equation \eqref{KS} one can 
see that the average of the temperature anisotropy values in the string 
component should go as the cosmic string tension squared. Therefore, if we 
define a reference cosmic string tension, $G\mu_0$, we have
\begin{equation}
\langle|T_S(k<k_p)|^2\rangle \, = \, 
\langle|T_S(k<k_p)|^2\rangle_0 \left(\frac{G\mu}{G\mu_0}\right)^2\,,
\end{equation}
where $\langle|T_S(k<k_p)|^2\rangle_0$ is the average for a string component 
corresponding to the reference tension and $G\mu$ is the cosmic string tension 
corresponding to the string component on the left-hand side of the equation. 
Substituting this into \eqref{power} we can solve for the final form of the 
scaling factor:
\begin{equation}\label{alpha}
\alpha^2 \, = \, 1 - 
\frac{\langle|T_S(k<k_p)|^2\rangle_0}{\langle|T_G(k<k_p)|^2\rangle}\left(\frac{G\mu}{G\mu_0}\right)^2\,.
\end{equation}
The benefit of having $\alpha$ in this form is we need only calculate the 
ratio of averages once using the reference tension. After this we can 
calculate the value of the scaling factor with only the cosmic string tension 
used in the given simulation, $G\mu$. As mentioned in the Introduction, 
studies of combining string anisotropies and Gaussian anisotropies 
\cite{Pogosian:2003mz,Pogosian:2008am} have concluded that, on the basis
of the angular power spectrum of CMB anisotropies, a cosmic string 
contribution of less than $10\%$ of the observed CMB power on large scales 
cannot be ruled out in general. However, in calculating the angular power
spectrum, coherent features in position space such as the line discontinuities
induced by the Kaiser-Stebbins effect are washed out. Thus, we expect
that better limits on the string tension can be established by making
use of edge detection algorithms working in position space.

A third component which we must include in the final map is a simulation of 
instrumental noise. For simplicity, we simulate an instrumental noise 
component that is simply white noise with some given maximum amplitude in the 
temperature difference $\delta T_{N,max}$. If an instrumental noise component 
is included we do not need to perform any additional scaling of the initial
Gaussian component of the map since the instrumental noise does not
modify the actual sky map. Thus, the instrumental noise component is simply 
summed directly to the other components. Denoting the noise component by 
$T_N(x,y)$, we have 
\begin{equation}
T(x,y) \, \equiv \, T_G(x,y) + T_N(x,y)
\end{equation}
for a simulation without cosmic strings, or
\begin{equation}
T(x,y) \, \equiv \, \alpha\, T_G(x,y) + T_S(x,y) + T_N(x,y)
\end{equation}
for a simulation including cosmic strings.

\begin{figure}[t]
\begin{center}
\subfigure{\includegraphics[width=0.32\linewidth]{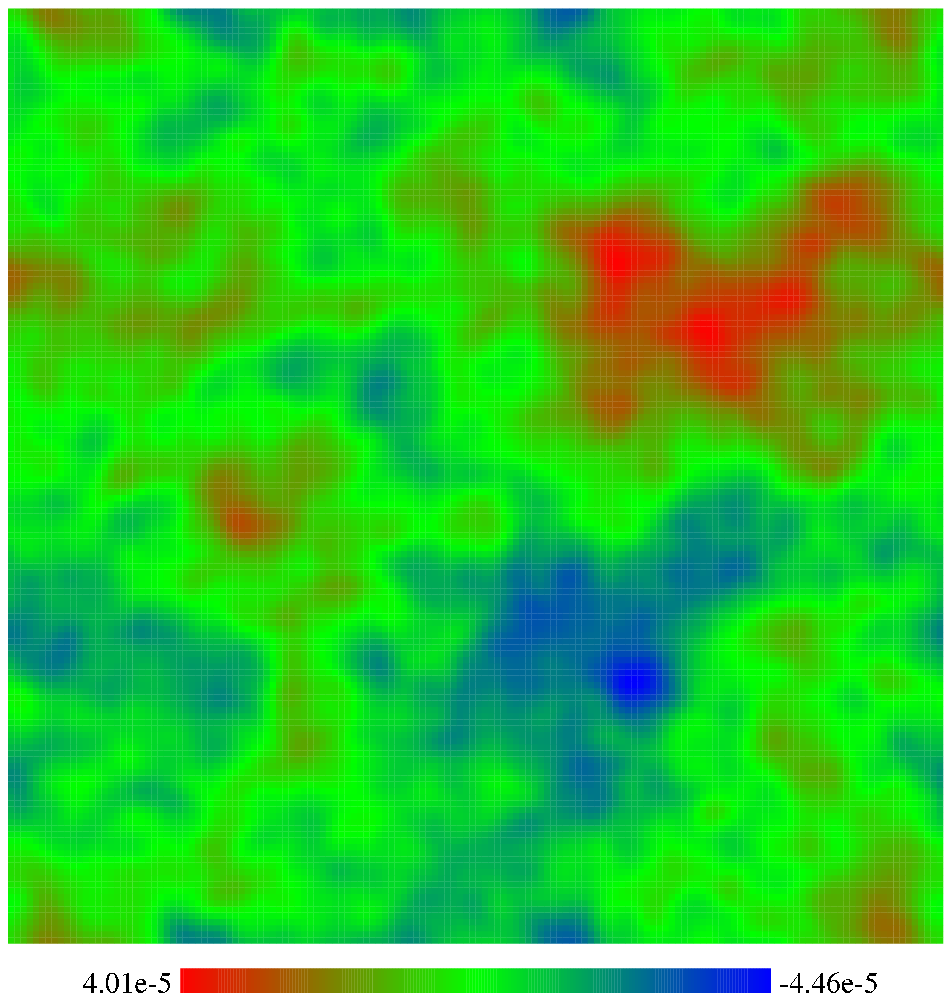}}
\subfigure{\includegraphics[width=0.32\linewidth]{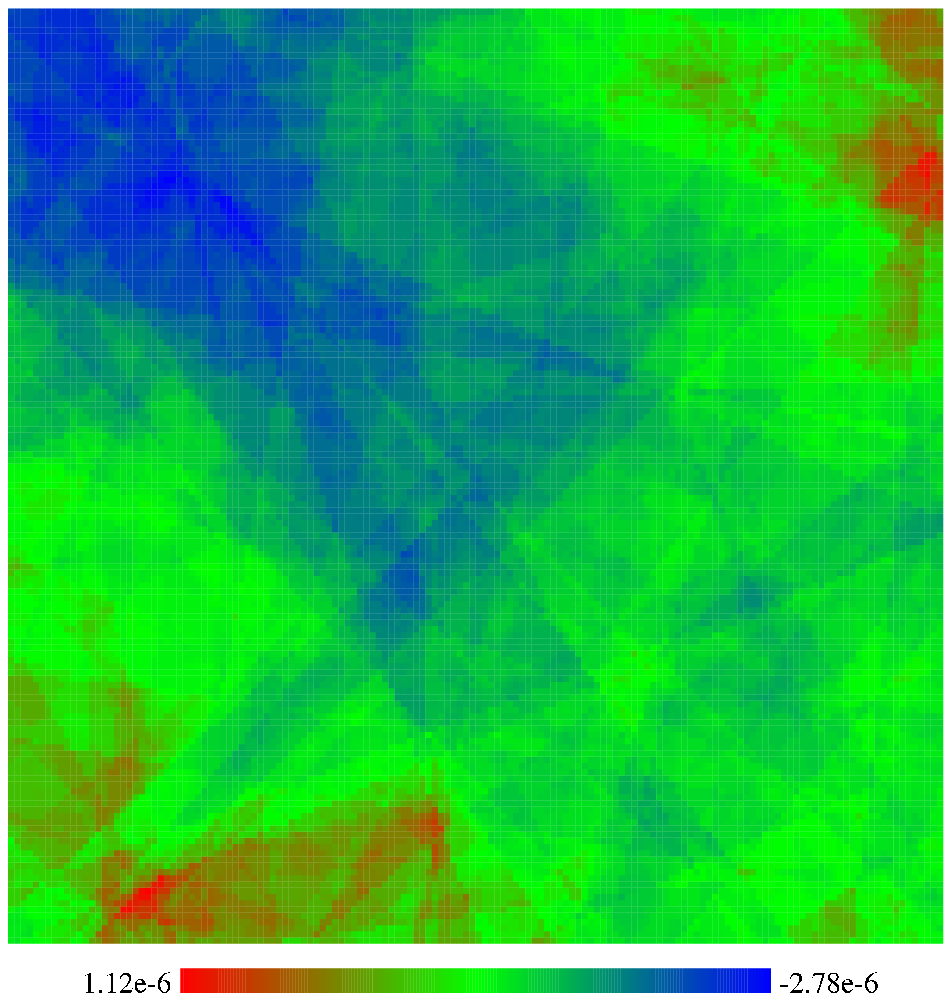}}
\subfigure{\includegraphics[width=0.32\linewidth]{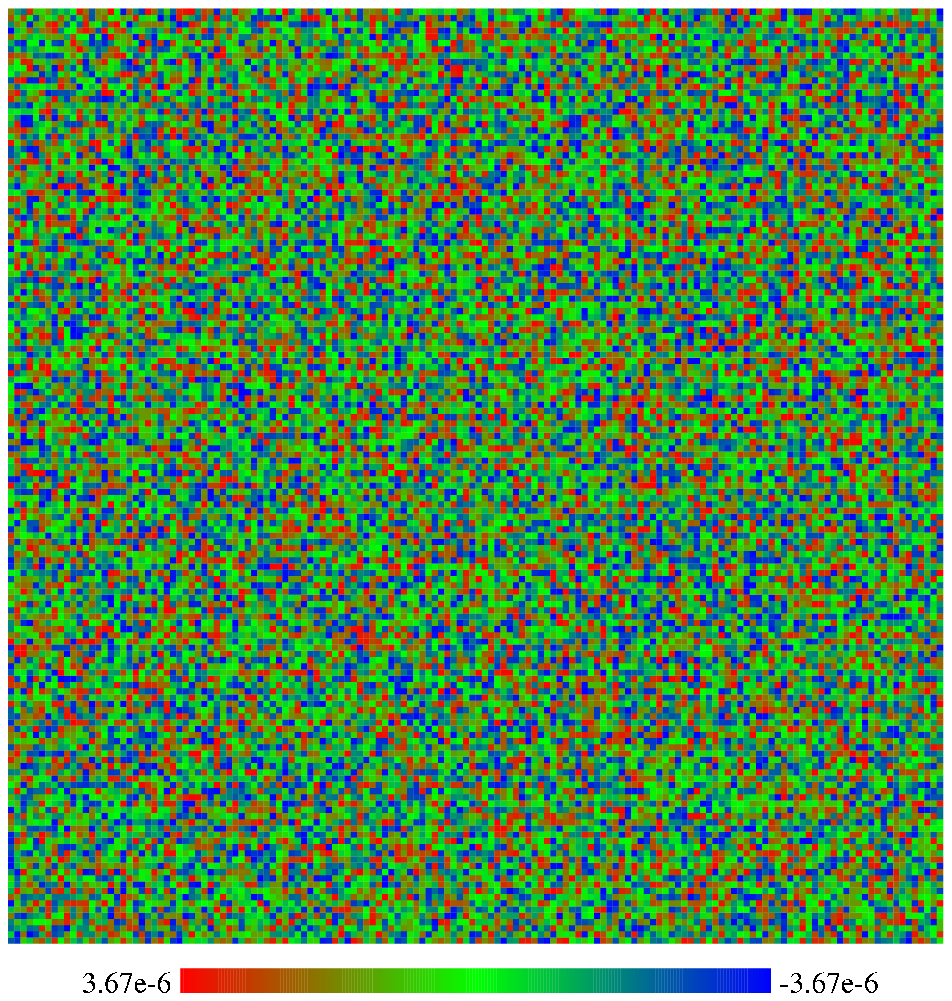}}
\caption{\label{comps} Components of a simulated temperature anisotropy map. 
On the left is a component of Gaussian temperature fluctuations, in the middle 
is a component of cosmic string induced temperature fluctuations and on the 
right is a component of instrumental noise. In all three components, the 
angular size of the simulated region is $2.5^\circ\times2.5^\circ$ and the 
angular resolution is $1^{\prime}$ per pixel (22,500 pixels). In the string 
component the tension of the cosmic strings was taken to be 
$G\mu=6\times10^{-8}$ and the number of strings per Hubble volume in the 
scaling solution was taken to be $M=10$. The colour of a pixel represents 
the value of the temperature anisotropy at that pixel, as described by the 
scale below each image.}
\end{center}
\end{figure}

The dominant portion of the final simulated map is the Gaussian temperature 
fluctuations. As such, these Gaussian fluctuations represent the most 
significant ``noise'' when trying to directly detect the effect of cosmic 
strings with the edge detection algorithm. The significance of the 
instrumental noise component in the final map is determined by the maximum 
amplitude of the noise, which should in general be small compared to the 
amplitude of the Gaussian fluctuations. The size of the temperature 
anisotropies in the string component depends directly on the tension of the 
cosmic strings which are being simulated, as described by Equation \eqref{KS}. 
For interesting values of the string tension, the amplitude of the 
string-induced anisotropies will lie from a factor of a few up to orders of 
magnitude below the amplitude of the Gaussian temperature anisotropies, thus 
presenting the difficulty in directly detecting them. Examples of each of the 
three map components are shown in Figure \ref{comps}. Before moving on to 
discuss the edge detection algorithm itself, we first review our methods for 
generating the Gaussian and string components.

\subsection{The Gaussian Component}

As touched on above, the spherical harmonic expansion of the CMB temperature 
anisotropies can be replaced by a Fourier expansion when using the flat-sky 
approximation \cite{White:1997wq}. Therefore, when generating the component 
of Gaussian fluctuations, we choose to work on a grid in Fourier space where 
each pixel in the grid is indexed by the coordinates $(k_x,k_y)$, which are 
the components of the wavevector pointing to that pixel. The size and 
resolution of the grid still correspond to the two angular scales in the 
simulation. The advantage of being able to use a Fourier analysis is that it 
greatly simplifies the calculations, and the value of the temperature 
anisotropy at a particular pixel on the the grid is then given by the relation
\begin{equation}\label{gtemp}
\frac{\delta T_{G}}{T}(k_x,k_y) \, = \, g(k_x,k_y)\,a(k_x,k_y)\,,
\end{equation}
where $g(k_x,k_y)$ is a random number taken from a normal probability 
distribution with mean zero and variance one \cite{Bond:1987ub}. The quantity 
$a(k_x,k_y)$ is the Fourier space equivalent of $a_{lm}$ in the usual 
spherical harmonic expansion and is related to the angular power spectrum of 
the temperature anisotropies in the same way,
\begin{equation}\label{a}
<|a(k_x,k_y)|^2> \, = \, C_l\,.
\end{equation}
In the flat-sky approximation the multipole moment is related to pixel 
position in the grid by
\begin{equation}\label{mpol}
l \, = \, \frac{2\pi}{\theta}\sqrt{k_x^2+k_y^2}\,,
\end{equation}
where $\theta$ is the angular size of the survey area \cite{Bond:1987ub}.

We compute the Fourier temperature fluctuations pixel by pixel using the 
above equations. It is clear from Equation \eqref{mpol} that, in general, the 
largest multipole moment required for a simulation increases as the resolution 
increases. Since we are interested in simulating high resolution CMB maps, we 
generate the COBE normalized angular power spectrum of the CMB to very large 
multipole moments using the \textsc{camb} software with input cosmological 
parameters determined by surveys at lower angular resolution. To be precise, 
we choose our input parameters to be those derived using the CMBall data set, 
which combines the results from multiple surveys \cite{Reichardt:2008ay}. 
Depending on the pixel position, the value of $l$ as calculated by Equation 
\eqref{mpol} can take non-integer values, whereas the angular power spectrum 
is computed for only integer values. In these cases, we simply approximate the 
value of the angular power spectrum at any given $l$ using a linear 
interpolation. Once we have computed the value of each pixel in the grid, we 
take the inverse Fourier transform of the array using a fast Fourier transform 
(FFT) algorithm, which produces a temperature anisotropy map in position space.

By choosing the origin of the grid to be at the top left corner in the maps, 
we have introduced a preferred direction into the simulation of the Gaussian 
fluctuations. To compensate for this asymmetry, we construct the final 
Gaussian component, $T_G(x,y)$, by superimposing four separate sub-components, 
which we label as $T_1...T_4$, each computed separately using the method 
described above. When combining these sub-components, we reflect each along 
one of the four axes on the grid eliminating any irregularity in the final 
map. Therefore, the Gaussian component is defined as
\begin{eqnarray}
T_G(x,y) \, &\equiv& \, \frac{1}{2}\big[T_1(x,y) + T_2(x_{max}-x,y) 
\nonumber \\ 
&& \, + T_3(x,y_{max}-y) + T_4(x_{max}-x,y_{max}-y)\big]\,,
\end{eqnarray}
where $x_{max}$ and $y_{max}$ are the maximal $x$ and $y$ values based on the 
simulation parameters. The factor of $1/2$ in front of the sum is required to 
maintain the original standard deviation.

\subsection{The String Component}

Since the focus of this work is on testing the edge detection method, not the 
details of the cosmic string network evolution, we utilize a toy model of the 
network for simplicity. We then examine the resulting temperature anisotropies 
caused by the strings which photons encounter between the time of last 
scattering and the present day. We choose to use the toy model originally 
presented by Perivolaropoulos in \cite{Perivolaropoulos:1992if}. 

In this model, we first separate the period between the present time, $t_0$, 
and the time of last scattering, $t_{ls}$, into N Hubble time steps such that 
$t_{i+1}=2t_i$. For a redshift of last scattering $z_{ls}=1000$ we then have 
\cite{Moessner:1993za}
\begin{equation}
N \, = \, \log_2\left(\frac{t_0}{t_{ls}}\right)\simeq15\,. 
\end{equation}
For large redshifts and assuming $\Omega_0=1$, the angular size of the Hubble 
volume at a given Hubble time is approximated by 
$\theta_{H_i}\sim z_{i}^{-1/2}\sim t_{i}^{1/3}$. Therefore, we have 
$\theta_{H_{ls}}\simeq z_{ls}^{-1/2}\simeq1.8^\circ$ for the Hubble volume 
corresponding to the time of last scattering and  
$\theta_{H_{i+1}}\simeq2^{1/3}\theta_{H_{i}}$ for all subsequent Hubble time 
steps \cite{Moessner:1993za}. At each Hubble time, a network of long straight 
strings with a length equal to two times the size of the Hubble volume at that 
time, each with random position, orientation and velocity, is laid down. The 
network of strings produced at each Hubble time is assumed to be uncorrelated 
with that of the previous Hubble time. This is justified since cosmic 
strings move with relativistic speeds, meaning that between Hubble times there 
will be multiple string interactions, causing the network to enter into a 
completely different configuration.

For a specific Hubble time step $t_i$ we start with an extended region that 
has a total angular size equivalent to the angular size of the string 
component being simulated plus two times the angular size of the Hubble volume 
at that particular Hubble time. The number of strings $n_i$ that should exist 
in that particular region is then given by the scaling solution
\begin{equation}
n_i \, = \, M\frac{(\theta+2\theta_{H_i})^2}{\theta_{H_i}^2}\,,
\end{equation}
where $M$ is the number of cosmic strings crossing each Hubble volume and 
$\theta$ is the angular size of the string component being simulated 
\cite{Moessner:1993za}. As usual, we work on a square grid, this time placed 
over the entire extended region, with pixel size still given by the angular 
resolution being considered. Pixels within the entire extended area are then 
chosen at random to be the midpoints of strings, with a probability such that 
the average number of strings in a single Hubble volume is in agreement with 
the number $M$ of the scaling solution. If a pixel is chosen to be a midpoint, 
we choose a random orientation about that pixel and we place a straight string 
of length $2\theta_{H_i}$. We then simulate the temperature fluctuation 
produced by that string by adding a temperature anisotropy
\begin{equation}\label{beta}
\frac{\delta T_S}{T} \, = \, 4\pi G\mu\gamma_s v_s r
\end{equation}
to a rectangular region on one side of the string, and subtracting the same 
amount from a rectangular region on the other side. This temperature 
anisotropy corresponds to the KS-effect as given by Equation \eqref{KS}, where 
$r=|\hat{k}\cdot(\hat{v}_s\times\hat{e}_s)|$ takes into account the projection 
effects. The direction of observation $\hat{k}$ is approximately constant over 
the entire field of view while the quantity $\hat{v}_s\times\hat{e}_s$ is a 
random unit vector since both the string orientation and velocity are random. 
Thus, the value of $r$ is uniformly distributed over the interval [0,1] 
\cite{Moessner:1993za}. In Equation \eqref{beta}, we take the RMS speed of the 
strings to be $v_s=0.15$ \cite{Moessner:1993za}, so the amplitude of the 
fluctuation is determined entirely by the string's tension and its 
orientation. Each rectangular region affected by the temperature fluctuation 
has a length $2\theta_{H_i}$ along the direction of the string and extends a 
distance $\theta_{H_i}$ in the direction perpendicular to the string 
\cite{Joao}. Thus, each cosmic string gives rise to five separate temperature 
discontinuities: one at its position, two parallel to it at a distance 
$\theta_{H_i}$ and two perpendicular to the string at the endpoints. After 
placing all of the cosmic strings and calculating the temperature fluctuation 
for each, we have finished simulating the cosmic string network for the given 
time step.

Since we began with a region which is larger than the string component we 
wanted to simulate in the first place, we must crop the larger area to the 
correct size. We choose to discard pixels equally from all four sides of the 
extended area, so that we retain only those from the central region of the 
larger area. By identifying the correctly sized simulated area with the centre 
of the extended area, one can see that what we essentially did when first 
defining the extended region was to enlarge the actual simulation area by a 
Hubble volume in each direction. The reason that we expand our simulated area 
in this way is because any string whose midpoint is within a distance 
$\theta_{H_i}$ of the actual area we want to simulate could enter into it. 
Thus, we must also account for these strings which lie around the edges of 
the area of interest, not only those centred within it. 

The final string-induced anisotropy map is given by the superposition of the 
effects of all of the strings in all of the Hubble volumes. Therefore, to 
produce the final cosmic string component $T_S(x,y)$, we simply sum together 
all fifteen sub-components pixel by pixel. This superposition approximates the 
contribution from the entire, more complex cosmic string network.

In the model described above, we have fixed values for the the speed of the 
strings, the length of the strings and the depth of the rectangular 
temperature fluctuation region around the string. These values were obtained 
from particular numerical simulations \cite{Moessner:1993za}, however, these 
parameters can vary significantly for different models of the string network 
(see \cite{2000csot.bookV} for a review) and should not be considered as 
established. We also note that in this toy model cosmic string loops and their 
subsequent effects are not included. Cosmic string loops will also
produce CMB anisotropies. However, based on the current knowledge of the
distribution of scaling strings, the string loop contribution to the CMB
is believed to be sub-dominant. This justified us neglecting these effects.

\section{The Canny Edge Detection Algorithm}
\label{seccanny}

\begin{figure}
\begin{center}
\subfigure{\includegraphics[width=0.32\linewidth]{smap_2.5x2.5.eps}}
\subfigure{\includegraphics[width=0.32\linewidth]{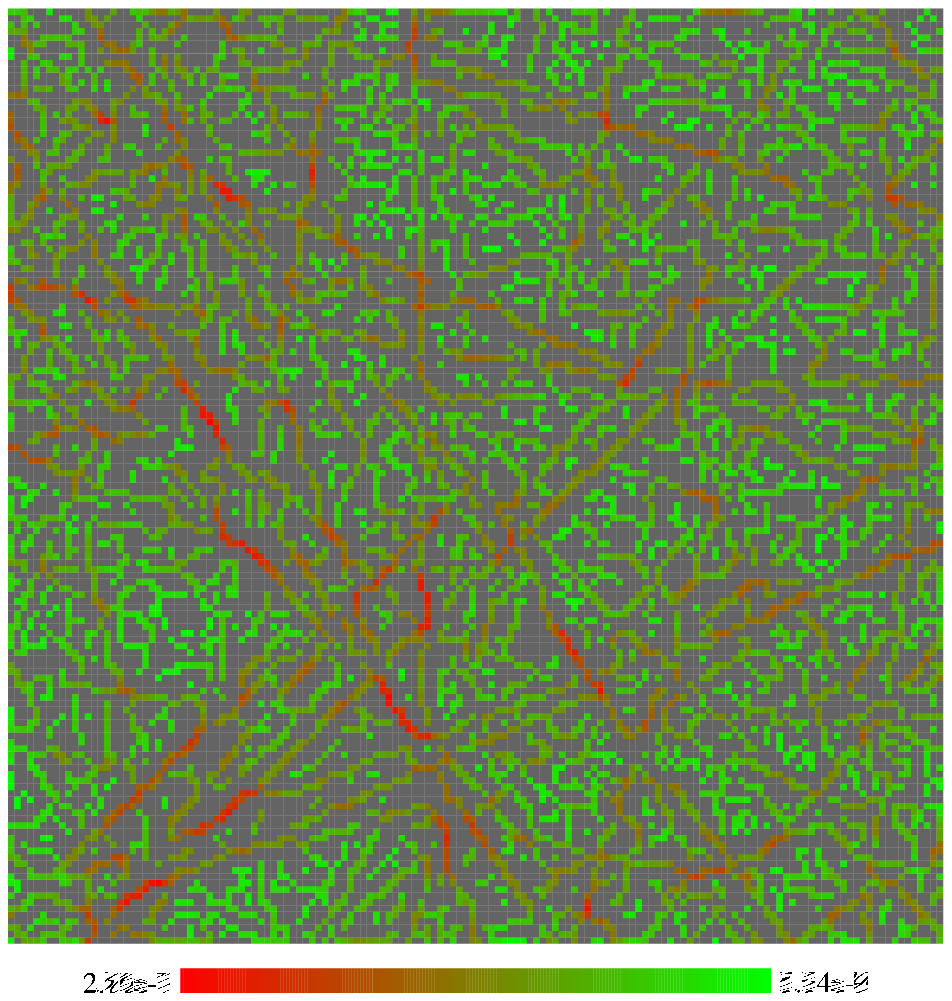}}
\subfigure{\includegraphics[width=0.32\linewidth]{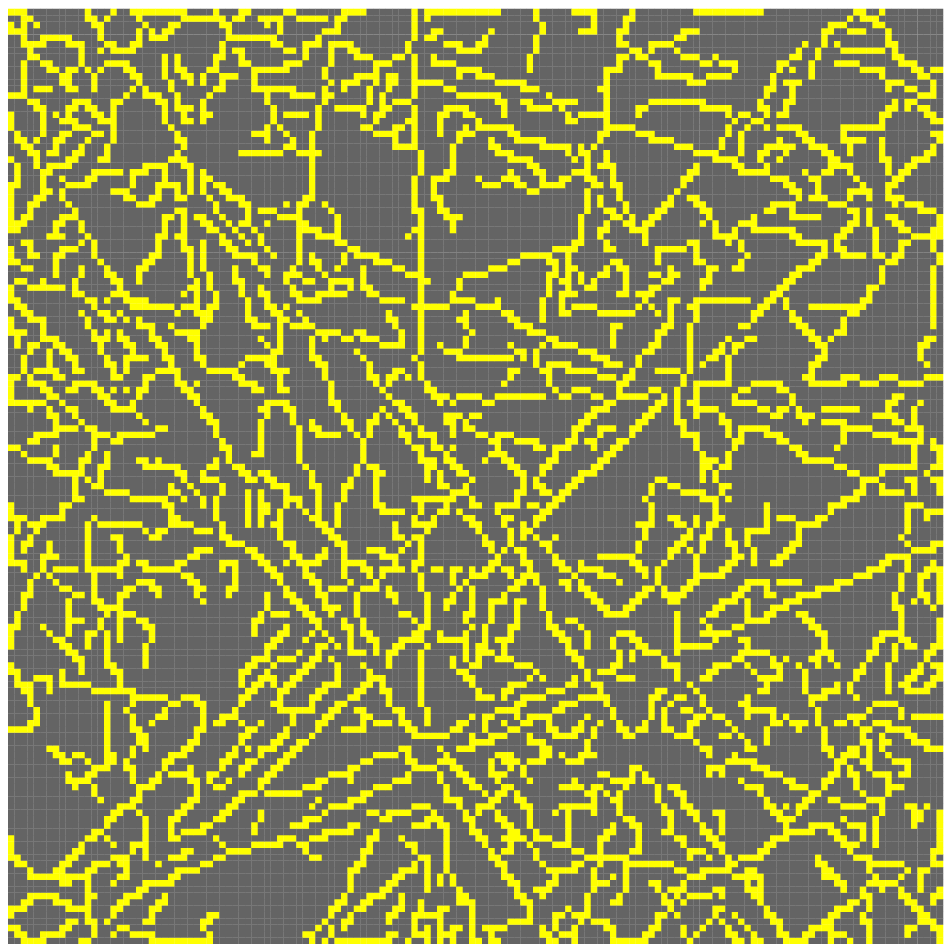}}
\caption{\label{canny} Maps produced by the Canny edge detection algorithm. 
On the left is a temperature anisotropy map that consists of cosmic string 
induced fluctuations only. The colour of a pixel represents the magnitude of 
the temperature anisotropy as described by the scale below the image. In the 
middle is an example of a map of local maxima generated after non-maximum 
suppression. The size of the gradient filters used was $5\times5$ pixels. The 
colour of a pixel represents the magnitude of the gradient at that pixel, as 
described by the scale below the image. On the right is an example of a final 
map of edges generated after thresholding with hysteresis. The values of the 
thresholds used were $t_u=0.25$, $t_l=0.10$ and $t_c=3.5$. The value of $G_m$ 
was calculated using a cosmic string tension of $G\mu<6\times10^{-8}$. The 
yellow pixels represent pixels which were determined to be on an edge. 
Together, these pixels show the the position, length and shape of the edges 
occurring in the original temperature anisotropy map. In last two maps, the 
grey pixels represent pixels which were discarded from the image.}
\end{center}
\end{figure}

When looking for edges in an image we are looking for curves across which 
there is a strong intensity contrast. The strength of an edge can then be 
quantified by the magnitude of the contrast from one side of the edge to the 
other, or equivalently, the magnitude of the gradient across the edge. For 
CMB temperature anisotropy maps, the intensity that we are dealing with is 
simply the amplitude of the fluctuations. Thus, we define the edges in the 
CMB maps as lines across which the temperature difference is large. To search 
for these edges we employ the Canny edge detection algorithm 
\cite{Canny:1986aa}, which is one of the most commonly used edge detection 
methods in image analysis. Figure \ref{canny} shows an example of a final map 
of edges generated by the Canny algorithm along with an intermediate map 
produced during the edge detection process. To clearly illustrate the result 
of each stage of the edge detection, we present maps corresponding to the same 
cosmic string component shown in Figure \ref{comps} with no other components 
added to it, however, this does not represent a legitimate final simulated CMB 
map. In the following sections we review the steps involved in applying the 
Canny algorithm to CMB maps and how these images are generated.

\subsection{Non-maximum Suppression}

Since we are interested in temperature gradients, the first step of the Canny 
edge detection algorithm is to simply compute the gradient of the temperature 
anisotropy map and use it to determine which pixels could be part of an edge. 
We first construct two square filters $F_x(x,y)$ and $F_y(x,y)$, which are 
first-order derivatives of a two dimensional Gaussian function along each of 
the two map coordinates $(x,y)$. We then apply each of these filters to the 
temperature map separately by convoluding the two using a FFT. This produces 
two new maps $G_x(x,y)$ and $G_y(x,y)$, which are the components of the 
gradient magnitude along the $x$-direction and $y$-direction. With these 
components we can then construct another new map
\begin{equation}
G(x,y) \, = \, \sqrt{G_x^2(x,y)+G_y^2(x,y)}\,,
\end{equation}
which is the map of the gradient magnitude, or edge strength, corresponding to 
the original temperature anisotropy map. We can also construct a second map
\begin{equation}\label{gang}
\theta_G(x,y) \, = \, \arctan\left(\frac{G_{y}(x,y)}{G_{x}(x,y)}\right)\,,
\end{equation}
which is the map of the gradient angle, or gradient direction. In the above 
equation the sign of both components is taken into account so that the angle 
is placed in the correct quadrant. Therefore, the arctangent has a range of 
($-180^\circ,180^\circ$]. However, at each pixel on a square grid there are 
only eight distinguished directions which form four axes. In order to relate 
the gradient direction as calculated by Equation \eqref{gang} to one that we 
can trace on the grid, we approximate the value of $\theta_G(x,y)$ at each 
pixel to lie along one of the eight grid directions. We do this by simply 
replacing the value of $\theta_G(x,y)$ with the angle corresponding to the 
closest grid direction. For example, if the gradient direction takes any of 
the values $-22.5^\circ\leq\theta_{G}(x,y)<22.5^\circ$ it would be replaced 
by $\theta_{G}(x,y)=0^\circ$.

In the Canny algorithm, part of the definition of a pixel that is considered 
to be on an edge is that it must be a local maximum in the gradient magnitude. 
By local maximum we mean that the gradient magnitude at a given pixel is 
larger than that of both pixels which neighbour it along the axis defined by 
the gradient direction at that same pixel. Using the gradient magnitude and 
direction maps, it is straightforward to check the local maximum condition 
pixel by pixel and determine which could be a part of an edge and which could 
not be part of an edge. Since we are only interested in constructing a final 
map of edges, if a pixel does not satisfy the local maximum condition we 
immediately discard that pixel. Therefore, this process is referred to as 
\emph{non-maximum suppression}.

Figure \ref{canny} shows a gradient magnitude map after non-maximum 
suppression has been performed. Many of the original pixels have been 
discarded, as expected, and we are left with a rough map of edges. Although 
curves corresponding to certain edges in the original temperature anisotropy 
component can be seen, there are many other pixels marked as local maxima 
corresponding to extremely weak edges, making the signal from stronger edges 
difficult to detect.

\subsection{Thresholding with Hysteresis}
\label{hysteresis}

When performing non-maximum suppression we only compared a single pixel with 
two of its neighbours to determine if it could be part of an edge. Pixels with 
a small gradient magnitude may have still been marked as a local maxima if the 
gradient magnitudes of their neighbours were also small. As mentioned above, 
Figure \ref{canny} shows that this is indeed the case. The magnitude at such 
pixels can in fact be so small that we do not want to consider them as edge 
pixels, since they can dilute the more significant signal coming from stronger 
edges. In addition, we want to detect edges which appear due to cosmic strings 
via the KS-effect. Therefore, we expect the gradient direction to be 
consistent across the length of the string induced edge. This directionality 
needs to be taken into account to determine which local maxima pixels belong 
to the same string edge. Taking these two points into consideration, we must 
further expand our definition of exactly what constitutes an edge pixel.

The Canny algorithm outlines a process of applying multiple thresholds to 
define the edges in an image, known as \emph{thresholding with hysteresis}. 
First, we choose an upper gradient threshold, $t_u<1$, such that we can then 
define a pixel which is definitely part of an edge, which we name a 
\emph{true-edge pixel}, as one which is not only a local maximum but also 
satisfies
\begin{equation}
G(x,y) \, \geq \, t_uG_m\,.
\end{equation}
Here $G_m$ is the mean maximum gradient magnitude computed from simulated 
temperature maps which contain only strings. The value of $G_m$ depends on the 
parameters of the simulation being performed, most notably the string tension, 
and must be computed separately for each parameter set using a selected number 
of simulated string maps. One can think of $G_m$ as representing the strongest 
possible edge that could be formed by cosmic strings alone. Therefore, with 
this threshold, we are simply stating that if the gradient magnitude at a 
given pixel is some chosen fraction of the maximum possible, then it must be a 
true-edge pixel. 

It is not sufficient, however, to define the edges using only one threshold 
because the gradient magnitude can fluctuate at each pixel along the length of 
an edge. This variation can be caused by both instrumental noise and the 
random nature of the Gaussian anisotropies. If we applied only an upper 
threshold, we would reject the pixels at which the gradient magnitude 
fluctuates below that threshold, but should in fact still be considered as a 
part of a given edge. This would lead to edges being cut into smaller 
segments, making them look like dashed lines, rather than continuous curves on 
the map. To avoid this, we also choose a lower gradient threshold, $t_l<t_u$, 
and define a pixel which is possibly part of an edge, which we name a 
\emph{semi-edge pixel}, as a local maximum pixel satisfying
\begin{equation}
t_lG_m \, \leq \, G(x,y) \, < \, t_uG_m\,.
\end{equation}
If a local maximum pixel still falls below the lower threshold then it is 
immediately rejected. The latter case is the requirement that an edge pixel 
have some minimum strength, and cures the problem of a local maxima with 
extremely small gradient magnitudes being included in the final edge map.

Since we are interested in edges appearing due of the presence of cosmic 
strings, we also apply a ``cutoff'' threshold such that we reject all pixels 
for which
\begin{equation}
G(x,y) \, > \, t_c G_m\,,
\end{equation}
where $t_c\geq1$. We apply this third threshold because the Gaussian 
temperature fluctuations in the CMB map dominate those coming from the cosmic 
strings. As such, they lead to edges with much stronger gradient magnitudes, 
that is, greater than $G_m$. If we only applied the upper bound $t_u$, these 
edges would overwhelm the edge detection algorithm, washing out the cosmic 
string signal. By setting a cutoff threshold, we can discard the pixels with a 
gradient magnitude which we consider to be too strong to have been caused by 
cosmic strings, and keep only those representing the cosmic string signature. 
We choose $t_c\geq1$ because we also consider the slight enhancement of weak 
edges corresponding to Gaussian fluctuations, as a result of the underlying 
cosmic string edges, to be part of the cosmic string signal.

After applying the thresholds as described above, we then further assert that 
any semi-edge pixel which is in contact with a true-edge pixel and has the 
appropriate gradient directionality is also a true-edge pixel sharing the same 
edge. By in contact, we mean that it is a semi-edge pixel which is one of the 
six neighbouring pixels of the true-edge pixel which does not lie along the 
gradient direction calculated at the position of the true-edge pixel. This 
definition stems from the fact that the two directions perpendicular to the 
gradient axis represent the edge axis, while the remaining four directions 
represent the two axes which are next to parallel to the edge axis. 
Essentially, we are stating that in order to be considered part of the same 
edge the semi-edge pixel must lie along (or almost along) the edge axis and it 
would be inconsistent for a pixel sharing the same edge to lie along the 
gradient direction. By appropriate gradient directionality, we mean that the 
semi-edge pixel also has a gradient direction which is parallel or next to 
parallel to the gradient direction calculated at the position of the true-edge 
pixel. The comparison of the gradient directions represents our demand that 
the temperature gradient be consistent along an entire edge.

We scan the remaining pixels in the map to check which semi-edge pixels 
satisfy the above conditions. The ones which do are immediately changed to 
true-edge pixels. This allows us to fill the gaps which occur between
true-edge pixels due to both types of noise, and avoid the incorrect breaking 
up of edges. Once a semi-edge pixel has been changed to a true-edge pixel it 
may then have another semi-edge pixel neighbouring it which needs to be 
changed, and so on. The scanning technique takes this into account and any 
connected series of semi-edge pixels will all be correctly changed to 
true-edge pixels ensuring that the entire edge is correctly identified. 
After scanning the map we consider all of the edges in the map to have been 
traced. At this point, if a pixel is still marked as a semi-edge pixel, we 
assume that it is not in contact with a true-edge pixel in any way, and it is 
rejected. The edge detection process is then finished, and the end result is 
the final map of true-edge pixels corresponding to the original temperature 
anisotropy map.

Figure \ref{canny} shows a final edge map after thresholding with hysteresis 
has been performed. Many of the pixels appearing in the map of local maxima 
have now been rejected, especially those with very small gradient magnitudes, 
and the stronger edges are now much better defined. This is a direct result of 
applying the thresholds and directionality conditions. Comparing the original 
temperature anisotropy map to the final edge map, it is clear that not only is 
the Canny algorithm good at locating the edges which are clearly visible, but 
that it is also sensitive to the faint edges which are not easily detectable 
by eye.

\section{Edge Length Counting and Statistical Analysis}
\label{seccount}

To facilitate a comparison with edge maps generated from different input 
temperature anisotropy maps, we need a way to quantify each individual edge 
map. Since we are considering cosmic strings as a source of edges in CMB 
temperature anisotropy maps, one might intuitively expect that in the presence 
of strings one would observe a larger number of edges of all lengths, or at 
least a larger number in some finite range of lengths. With this in mind, we 
employ a simple method of quantifying the edge maps, which is to record the 
length of each edge appearing in the edge map. 

We define a single edge as a chain of true-edge pixels where each subsequent 
pixel is in contact with the previous pixel and has a similar gradient 
direction (both in the same sense as described in the previous section). When 
scanning the final edge map, we count the number of pixels appearing in each 
separate edge. These values are exactly the lengths of the edge in units of 
pixels. With this data we can then construct a histogram of the total number 
of edges of each possible length, which corresponds to the original 
temperature anisotropy map. We do not consider a single pixel to represent an 
edge, therefore, the minimum edge length that we include in our histograms is 
two pixels long. If there are any single pixels marked as edges then we simply 
ignore them.

After generating histograms for different input maps, we need to develop a way 
to compare them and look for differences. Specifically, we are looking for a 
change in the distribution of the total number of edges between an edge map 
corresponding to a simulation without cosmic strings and an edge map 
corresponding to a simulation with cosmic strings. However, both the Gaussian 
and string components in the simulated CMB temperature anisotropy maps are 
generated using random processes. If we were to compare two histograms 
generated from only one simulated temperature anisotropy map each, we would 
not be able to draw a very meaningful conclusion. Therefore, to make our 
comparison more robust, we simulate many temperature anisotropy maps with the 
same input parameters and perform the edge detection and length counting on 
each one separately. This provides a set of histograms from which we can then 
compute the mean number of edges of each length occurring over all the runs. 
We also compute the standard deviation from each mean value. In the end this 
provides us with a new \emph{averaged histogram} of edge lengths that has 
statistical error bars. Comparing two of these averaged histograms then allows 
us to assign a statistical significance to the difference in the 
distributions. From this point on, whenever we mention a histogram we mean an 
averaged histogram computed using many simulations.

When comparing two histograms, we compare the mean value for each specific 
length separately, rather than perform a single general test based on the 
overall shapes of the distributions. We prefer to treat each bin separately 
because each has a separate standard deviation associated with it. 
Furthermore, we assume that the underlying values used to compute each mean 
are normally distributed. That way we can use Student's t-statistic to 
determine the significance of the difference at each length.

For two samples of equal size $n$, Student's t-statistic is defined as
\begin{equation}\label{t}
t \, = \, \left(\overline{N}_1-\overline{N}_2\right)\sqrt{\frac{n}{(\sigma_{1})^{2}+(\sigma_{2})^2}}\,,
\end{equation}
where $\overline{N}_1$ and $\sigma_{1}$ are the mean and sample standard 
deviation of the first sample and $\overline{N}_2$ and $\sigma_{2}$ are the 
mean and sample standard deviation of the second sample. Given two histograms, 
we compute $t$ for each length occurring in the two histograms for which 
$\overline{N}_i\geq3\sigma_i$, where $i=1,2$. This constraint on the lengths 
we consider stems from our assumption that the underlying distribution of each 
mean value is normal. Since it would be inconsistent to consider negative 
values for the total number of strings at any given length, we choose only 
lengths for which the total number of strings is positive definite at the 
$3\sigma$ level. The p-value corresponding to each $t$ is then computed from 
a t-distribution with $2n-2$ degrees of freedom.

We then combine the probabilities calculated for each length, denoted by 
$p_L$, into a single statistic which characterizes the difference between the 
two histograms. Using Fisher's combined probability test, we can define the 
new statistic $\chi^2$ as
\begin{equation}
\chi^{2} \, = \, -2\sum^{L_{m}}_{L=2}\ln(p_{L})\,,
\end{equation}
where $L_{m}$ is the maximum length at which a p-value was computed. The final 
p-value corresponding to the statistic $\chi^2$ is then determined from a 
chi-square distribution with $2L_{m}-2$ degrees of freedom.

The final step is to compare this single p-value to a significance level 
$\epsilon$ to conclude whether or not the difference in the two histograms is 
significant. We choose to work with the customary significance level 
$\epsilon=0.0027$ corresponding to $3\sigma$ of a normal distribution. If 
our p-value is less than $\epsilon$ we state that the difference in the two 
edge maps is statistically significant.

\section{Results}
\label{secresults}

We present the results of running the Canny edge detection algorithm on 
simulated CMB anisotropy maps in two parts. First, we report the results for 
simulations which are designed to mimic the expected output from the SPT. 
Using these results, we determine what kind of bound on the cosmic string 
tension one could hope to achieve using the edge detection method on data from 
that survey. Second, we present the results for simulations corresponding to a 
hypothetical survey that has different specifications than those of the SPT. 
We use these results to investigate how the potential constraint on the 
tension changes with respect to the design of the survey.

\begin{figure}
\begin{center}
\subfigure{\includegraphics[width=0.32\linewidth]{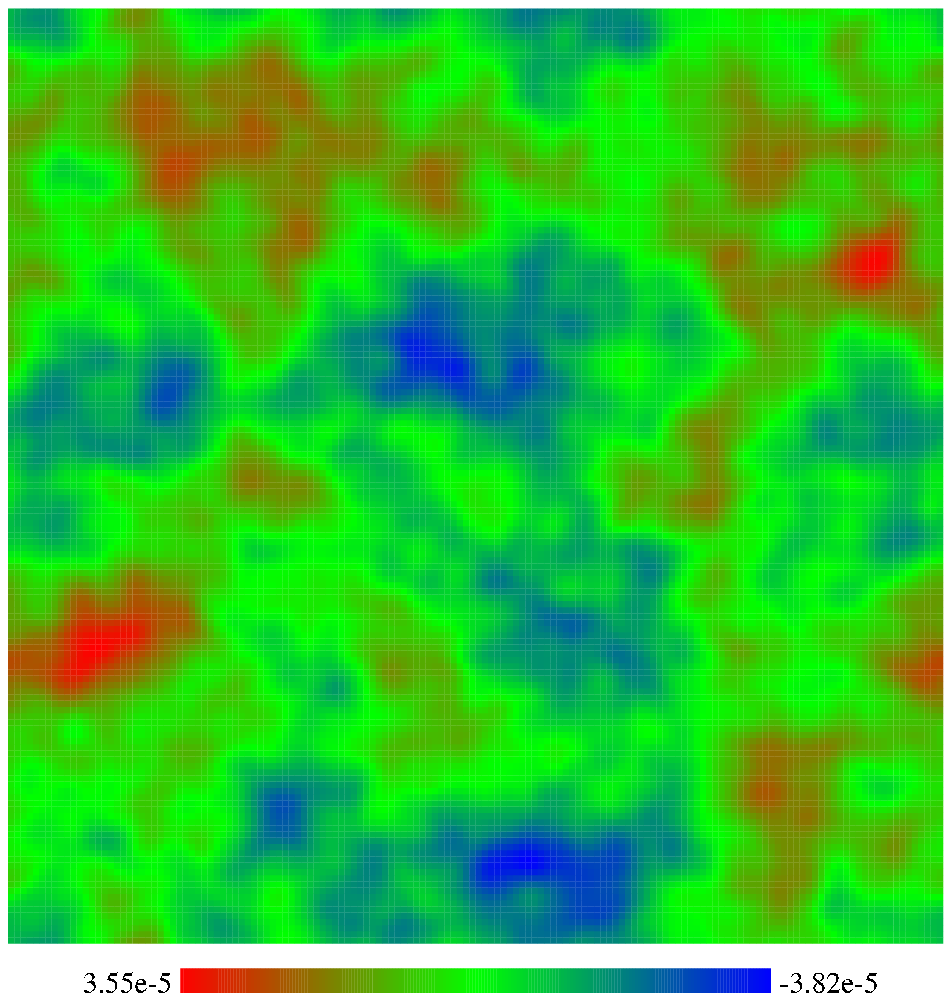}}
\subfigure{\includegraphics[width=0.32\linewidth]{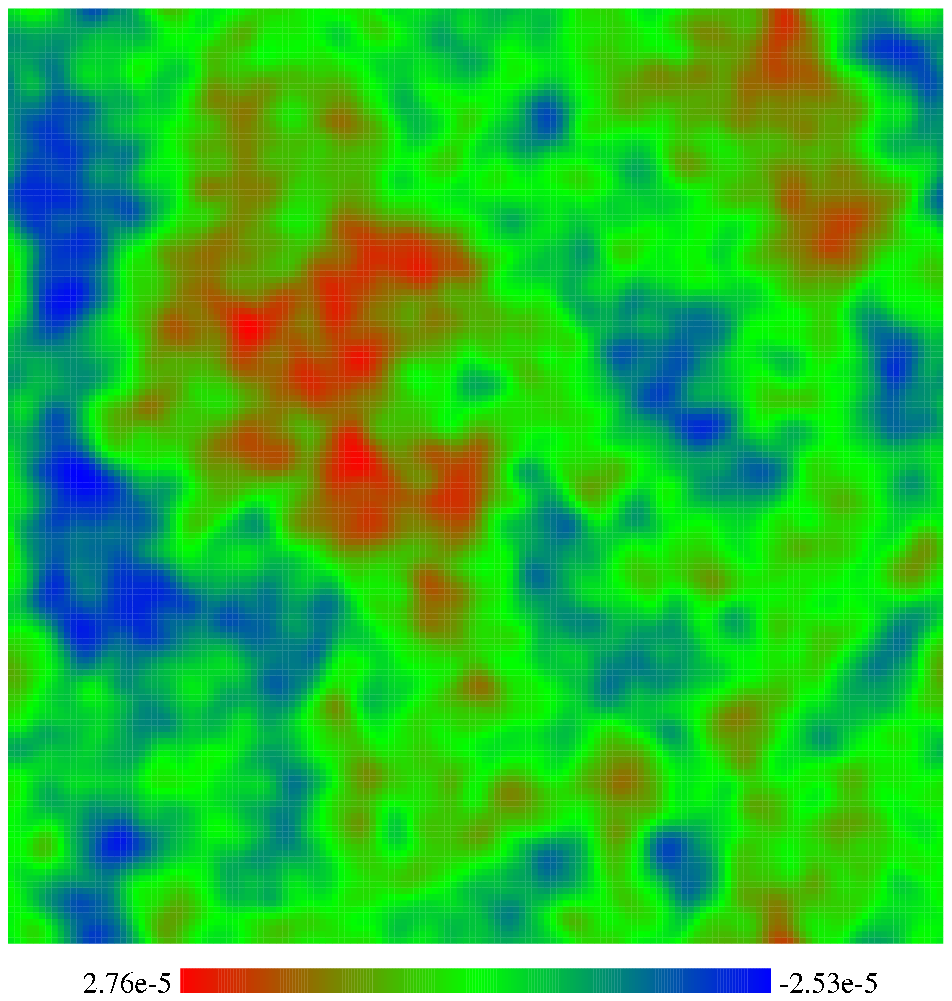}}
\subfigure{\includegraphics[width=0.32\linewidth]{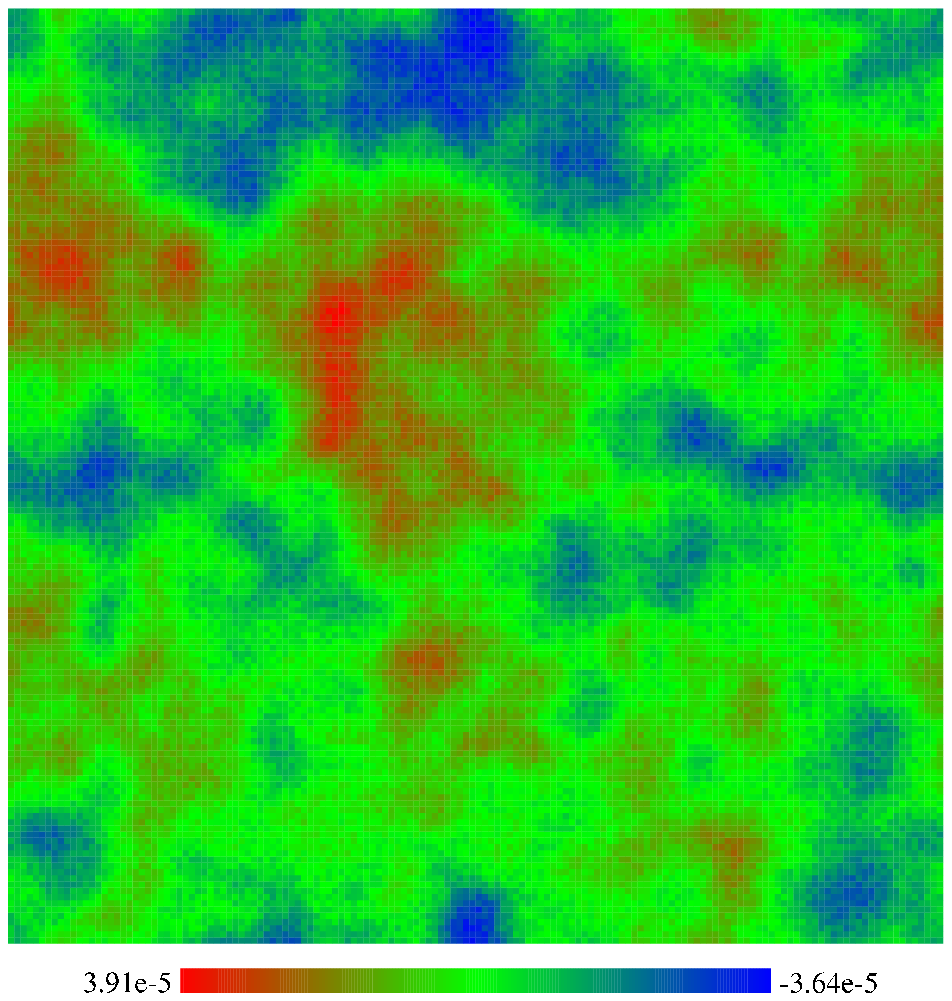}}
\caption{\label{compare} Comparison of CMB maps including different components.
On the left is a basic map map with Gaussian fluctuations only. In the middle 
is a map which includes a cosmic string component but excludes instrumental 
noise. On the right is a map which includes instrumental noise but excludes a 
cosmic string component. Note that different realizations of the Gaussian noise were used in each of the three panels. All three maps show a $2.5^\circ\times2.5^\circ$ 
patch of sky at 1$^\prime$ resolution (22,500 pixels). The values of the free 
parameters in the cosmic string simulation were $G\mu=6\times10^{-8}$ and 
$M=10$. The scaling factor in the map component addition that produced the 
map in the middle was $\alpha=0.987$. For the map on the right the maximum 
temperature fluctuation caused by the noise was taken to be 
$\delta T_{N,max}=10$\:$\mu$K.}
\end{center}
\end{figure}

\begin{figure}
\begin{center}
\includegraphics[width=0.75\linewidth]{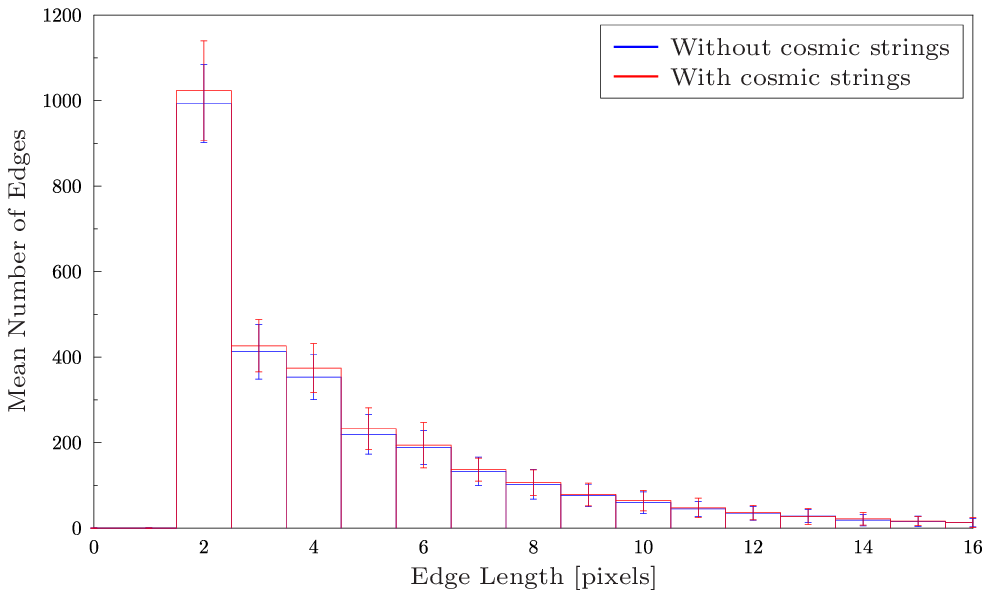}
\caption{\label{comparehist} Comparison of histograms for maps with and 
without a component of cosmic string induced fluctuations. Each histogram 
corresponds to a set of 40 simulated CMB maps. The angular size of each map 
was $10^\circ\times10^\circ$ and the angular resolution of each was 1$^\prime$ 
per pixel (360,000 pixels). In the maps including a cosmic string component, 
the string free parameters were taken to be $G\mu<6\times10^{-8}$ and $M=10$ 
while the scaling factor in the map component addition was $\alpha=0.987$. In 
the edge detection algorithm the gradient filter length was 5 pixels and the 
thresholds were $t_u=0.25$, $t_l=0.10$ and $t_c=3.5$. The value of $G_m$ was 
calculated using the same cosmic string tension given above. The height of 
each bar corresponds to the mean number of edges at that edge length. The 
error bars represent a spread of $3\sigma$ from the mean value, where $\sigma$ 
is the standard deviation of the mean. Shown here are only the lengths for 
which the mean is greater than $3\sigma$ in both histograms.}
\end{center}
\end{figure}

The SPT is capable of producing a 4,000 square degree survey of the 
anisotropies in the CMB \cite{Ruhl:2004kv}. To replicate the same amount of 
sky coverage, we simulate 40 separate $10^\circ\times10^\circ$ maps, where the 
angular resolution of each of these maps is 1$^\prime$ per pixel, again 
matching that specified for the SPT. To test the edge detection method, we 
simulate two separate sets of 40 maps, the first set including the effect of 
cosmic strings, and the second set excluding the effect of cosmic strings. 
Each set of maps gives rise to a histogram of edge lengths via the edge 
detection and edge length counting algorithms. We then compare these two 
histograms using the statistical analysis described in Section \ref{seccount} 
to determine if the difference in the distributions is significant. We repeat 
this process for many different values of the cosmic string tension, until we 
can no longer identify a statistically significant difference in the two 
histograms. Figure \ref{compare} shows a side by side comparison of a 
simulated CMB map without a cosmic string component and a simulated CMB map 
which does include a cosmic string component. The effect of the cosmic strings 
in the final temperature anisotropy map is not apparent and any difference in 
the typical structure between the two maps is unnoticeable by eye. Figure 
\ref{comparehist}, on the other hand, shows a histogram corresponding to a 
set of maps without a cosmic string component and a histogram corresponding to 
a set of maps with a cosmic string component. The two histograms show that the 
edge detection method is in fact able to detect a difference which is not 
evident by eye, with maps including strings having slightly higher mean values 
for certain lengths. Although the difference in histograms may not seem large, 
this particular example would generate a significant result.

Although the angular size and resolution of the simulation are determined by 
the specifications of the survey in question, the values of the other free 
parameters in each step of process must also be fixed. We take the number of 
cosmic strings per Hubble volume in all of the string component simulations to 
be $M=10$ \cite{Moessner:1993za}, regardless of the cosmic string tension. In 
every run of the edge detection algorithm, we choose the gradient filter 
length to be 5 pixels, the value of the upper threshold to be $t_u=0.25$ and 
the value of the lower threshold to be $t_l=0.10$. These values for the 
thresholds may appear small, but as one can see from the scale in Figure 
\ref{canny}, the gradient magnitude in the string component can take a large 
range of values. Therefore, $G_m$ can be quite a bit larger than the average 
gradient magnitude on a string induced edge, so we must choose low values for 
the thresholds in order to not throw away the entire string signal. We have 
not mentioned the value of the scaling factor in the map addition, $\alpha$, 
nor the value of the cutoff threshold, $t_c$. The reason is, we do not fix the 
value of these two parameters for all of the runs. In the case of the scaling 
factor, its value must change for each given cosmic string tension, as 
described by Equation \eqref{alpha}. The value of the cutoff threshold, on the 
other hand, is chosen deliberately based on the value of the tension, such 
that we get the best results from our edge detection method. We note the value 
of both of these parameters when presenting our findings.

For the SPT specific simulations, the capability of the edge detection method 
to make a significant detection of the cosmic string signal for different 
choices of the cosmic string tension is summarized in Table \ref{SPT}. We find 
that our edge detection method can distinguish a signal arising from cosmic 
strings down to a tension of $G\mu=5\times10^{-8}$. Therefore, if the edge 
detection method was used on ideal data from the SPT, but was unable to 
distinguish a difference from a theoretical data set without the effect 
cosmic strings, we could then impose a constraint on the cosmic string 
tension of $G\mu<5\times10^{-8}$.

\begin{table}[t]
\caption{\label{SPT} Summary of the ability of the Canny algorithm to make a 
significant detection of a cosmic string signal for SPT specific simulations. 
Shown here are the results corresponding to simulated CMB maps excluding 
instrumental noise as well as simulated CMB maps including instrumental noise. 
In the first column are different choices for the tension of the cosmic 
strings. In the second, third and fourth columns are the values of the scaling 
factor, cutoff threshold and p-value respectively, corresponding to each of 
the tensions. A p-value of less than $2.7\times10^{-3}$ indicates that the 
simulations including cosmic strings produced significantly different results 
from those without cosmic strings.}
\centering 
\begin{tabular}{c|c|c|c}
\hline\hline
String Tension ($G\mu$) & Scaling Factor ($\alpha$) & Cutoff Threshold ($t_c$) & p-value \\
\hline
\multicolumn{4}{c}{Without Instrumental Noise}\\
\hline
$6.0\times10^{-8}$ & 0.987 & 3.5 & $7.19\times10^{-12}$ \\
$5.5\times10^{-8}$ & 0.989 & 4.2 & $6.99\times10^{-4}$ \\
$5.0\times10^{-8}$ & 0.991 & 5.5 & $2.39\times10^{-3}$ \\
$4.5\times10^{-8}$ & 0.993 & 6.0 & $9.95\times10^{-3}$ \\
\hline
\multicolumn{4}{c}{With Instrumental Noise}\\
\hline
$6.0\times10^{-8}$ & 0.987 & 3.5 & $2.92\times10^{-10}$ \\
$5.5\times10^{-8}$ & 0.989 & 4.2 & $1.45\times10^{-3}$ \\
$5.0\times10^{-8}$ & 0.991 & 5.5 & $1.36\times10^{-2}$ \\
$4.5\times10^{-8}$ & 0.993 & 6.0 & $1.96\times10^{-2}$ \\
\hline
\end{tabular}
\end{table}

The above mentioned results were determined from simulated maps which did not 
contain a component of instrumental noise. To examine the effect that detector 
noise will have on the ability of the edge detection method to constrain the 
cosmic string tension, we repeat the same process described above, with the 
same choices for all of the parameters, but this time with instrumental noise 
included in the simulation of the CMB maps. As mentioned earlier, we simulate 
a component of white noise with a given maximum temperature change. Here, we 
choose the maximum temperature change caused by the instrumental noise to be 
$\delta T_{N,max}=10\:\mu\mbox{K}$, roughly corresponding to that planned for 
the SPT \cite{Ruhl:2004kv}. Figure \ref{compare} shows a side by side 
comparison of a simulated CMB map which includes instrumental noise and one 
which does not. The effect that the noise has on the map is clear, making it 
appear pixelated and non-Gaussian, yet the overall structure of the image is 
still visible since the temperature fluctuations caused by the noise are 
sub-dominant compared to the Gaussian fluctuations.

For the SPT specific simulations including instrumental noise, the results of 
using the edge detection method to detect a cosmic string signal are also 
presented in table \ref{SPT}. We find that detector noise does not have a 
substantial effect, and it weakens the possible constraint that the edge 
detection method could place on the cosmic string tension only slightly to 
$G\mu<5.5\times10^{-8}$.

\begin{table}[t]
\caption{\label{bigSPT} Summary of the ability of the Canny algorithm to make 
a significant detection of a cosmic string signal for simulations 
corresponding to a hypothetical CMB survey. Shown here are the results 
corresponding to simulated CMB maps excluding instrumental noise and simulated 
CMB maps including instrumental noise. See the caption of Table \ref{SPT} for 
a description of the columns.}
\centering 
\begin{tabular}{c|c|c|c}
\hline\hline
String Tension ($G\mu$) & Scaling Factor ($\alpha$) & Cutoff Threshold ($t_c$) & p-value \\
\hline
\multicolumn{4}{c}{Without Instrumental Noise}\\
\hline
$3.5\times10^{-8}$ & 0.995 & 8.0 & $1.95\times10^{-5}$ \\
$3.0\times10^{-8}$ & 0.997 & 8.8 & $8.16\times10^{-4}$ \\
$2.5\times10^{-8}$ & 0.998 & 9.6 & $7.87\times10^{-3}$ \\
\hline
\multicolumn{4}{c}{With Instrumental Noise}\\
\hline
$3.5\times10^{-8}$ & 0.995 & 8.0 & $2.37\times10^{-5}$\\
$3.0\times10^{-8}$ & 0.997 & 8.8 & $1.46\times10^{-3}$ \\
$2.5\times10^{-8}$ & 0.998 & 9.6 & $2.80\times10^{-1}$ \\
\hline
\end{tabular}
\end{table}

Along with the results specific to the SPT, we explore how the constraint 
which could be applied by the edge detection method changes based on the 
specifications of the survey. For this purpose, we imagine a theoretical 
observatory which has the same specifications as the SPT but could map five 
times the amount of sky with the same resolution, that is, produce a 20,000 
square degree survey of the anisotropies in the CMB. To replicate the output 
of a survey with this design, we instead simulate 200 separate 
$10^\circ\times10^\circ$ maps at $1^{\prime}$ resolution. In this hypothetical 
case we again choose the maximum temperature change caused by the instrumental 
noise to be $\delta T_{N,max}=10\:\mu\mbox{K}$. The analysis follows the same 
procedure as outlined above, and we keep the same values for all of the free 
parameters.

For the larger survey size, the results of using the edge detection method to 
detect a cosmic string signal are summarized in Table \ref{bigSPT}. By 
increasing the survey size from that of the SPT by a factor of five, while 
keeping all other specifications the same, the ideal output from such an 
observatory could have the potential to improve the constraint on the cosmic 
string tension to $G\mu<3.0\times10^{-8}$. When instrumental noise is 
included, we find that the effect on the edge detection method is in this case 
negligible and the possible bound remains the same as that found using 
simulations without instrumental noise.

\section{Discussion}
\label{secdiscuss}

We have developed a method of searching for linear discontinuities in the 
microwave background temperature caused by the presence of cosmic strings 
along our line of sight to the surface of last scattering. The method which 
we have developed involves applying an edge detection algorithm to CMB 
temperature anisotropy maps in order to identify the effect of cosmic strings. 
We have applied our edge detection method to simulated CMB maps both including 
cosmic strings, and without cosmic strings, to test its ability to 
discriminate between the two. This then translates directly into a possible 
constraint on the cosmic string tension. In particular, we have focused on 
two different sets of simulations, one which mimics the future output coming 
from the SPT and one which corresponds to a theoretical survey which covers 
five times as much sky as the SPT with the same angular resolution. We find 
that the edge detection method could potentially place a bound on the cosmic 
string tension of $G\mu<5\times10^{-8}$ for a perfect CMB observation from the 
SPT and that this could be lowered to $G\mu<3\times10^{-8}$ for the larger 
survey\footnote{Note that the dependence of the limit as a function of angular resolution was studied in \cite{Amsel:2007ki}}. For more realistic simulations which include instrumental noise, we 
find that the potential bound corresponding to the SPT weakens by only a small 
amount to $G\mu<5.5\times10^{-8}$ while the possible bound corresponding to 
the theoretical survey does not change at all, and is still 
$G\mu<3\times10^{-8}$. We consider the constraint corresponding to the SPT 
specific simulations which include a component of instrumental noise to be the 
main conclusion of this work. This possible bound is approximately an order of 
magnitude better than those arising from other methods which use CMB 
observations and approximately two orders of magnitude better than those 
arising from other methods which search for the KS-effect. Therefore, we 
believe that using the output from the SPT along with the edge detection 
method has the potential to greatly improve the constraint on the cosmic 
string tension. This bound is not tighter than the constraints arising from 
current pulsar timing data, although it is competitive, falling directly 
within the range of values reported by different observations. Nevertheless, 
as mentioned in the Overview, we consider our method of constraining the 
tension to be more robust since we make less assumptions about some of the 
unknown parameters which describe the cosmic string network and its evolution. 
Therefore, we believe that the possible bound on $G\mu$ given above would in 
fact represent a stronger constraint.

We conclude that instrumental noise does not have a very major effect on the 
ability of the edge detection method to identify the cosmic string signal. We believe that this is an indication that the 
thresholding with hysteresis performs as it should, since noisy pixels could 
destroy the edge signal by causing large fluctuations in the gradient 
magnitude. Furthermore, as one can see from Figure \ref{comparehist}, the 
largest difference between histograms occurs at short lengths rather than 
longer lengths. The instrumental noise leaves this difference in the short 
edge signal between maps with and without strings relatively unchanged, since 
the probability of a particularly noisy pixel falling on a short edge, 
resulting in it being incorrectly detected by the Canny algorithm, is small 
compared to that for longer edges. While on the topic of instrumental noise, we reiterate that we have included only a simplified white noise component in our simulations. A more complex investigation of instrumental noise would include a low frequency piece which results in stripes appearing in the final map of the CMB. Based on the method described in this paper, it is clear that striping would be crucial, since it would result in maps with more edges than that predicted by the cosmological theory and this could be confused with the effect of cosmic strings. One redeeming feature of this type of low frequency noise is that the stripes which are introduced would lie along the scanning direction, thus, when dealing with actual SPT data, it may be possible to subtract this effect out of the final map or to simply ignore edges lying along the known scanning direction in the edge detection algorithm itself. No other systematic effects due to the instrumental scanning strategy have been included in the current analysis, nor have errors due to foregrounds in the microwave sky. In future work it would be useful to investigate all types types of noise as well as the removal strategies in more detail to determine if they could change the behavior of the edge detection method.

We also found that increasing the simulated 
survey size increases the statistical significance of the deviations between 
the histograms for similar values of the cosmic string tension. This behaviour 
is expected though, since more edge maps were used to compute the mean values 
in each of the histograms and one can see from Equation \eqref{t} that the 
value of $t$ scales as $\sqrt{n}$. While the p-values are smaller for similar 
tensions, the final constraint which can be levied by the larger survey is not 
drastically different from that corresponding to the SPT specific simulations. 
Increasing the survey size by 5 times only lowered the possible constraint by 
a factor of roughly $\sqrt{5}$. Based on this result, we conclude that the 
survey size does not have a major influence on the ability of the edge 
detection method.

When generating the simulated CMB maps, we employed a toy model of the cosmic 
string network which includes only straight strings and no cosmic string 
loops. More detailed models of the network and its evolution have been 
developed in other works 
\cite{Albrecht:1989mk,PhysRevLett.60.257,Allen:1990tv,Fraisse:2007nu} and can 
be implemented numerically. Therefore, one obvious way to improve the testing 
method we have outlined here would be to implement one of these more complex 
models which would in turn produce a more realistic map of the temperature 
anisotropies induced via the KS-effect. On the other hand, we stress that a 
change of this nature would come at a large computational expense. On a 
similar note, it may also be useful to develop a more robust method of 
combining the string induced temperature anisotropies with those coming from 
Gaussian fluctuations, to make sure that the final simulated map agrees with 
other observations. Furthermore, after applying the Canny edge detection 
algorithm to the CMB temperature anisotropy maps, we quantify the 
corresponding edge map by recording the length of every edge appearing in it. 
As mentioned in Section \ref{seccount}, this is one of the simplest ways of 
describing the edge map, and it may be beneficial to investigate an 
alternative method of image comparison which provides a more powerful way of 
discriminating between the two edge maps. For now, we leave these improvements 
as the goal of future work. 

While we have chosen to focus on the SPT in this work, the edge detection 
method is quite versatile and could be used with virtually any high resolution 
CMB survey. We conclude that this method presents a powerful and unique way of 
constraining the cosmic string tension which has the potential to perform 
better than current methods, or, at the very least, to provide a complimentary 
technique to those already in use.

\section*{Acknowledgements}

We would like to thank Gil Holder and Matt Dobbs for useful discussions concerning 
various parts of this work. Thank you to Joshua Berger and, especially, 
Stephen Amsel for making their code available and for answering many questions 
regarding the edge detection method. We also wish to extend a big thank to Eric Thewalt for 
debugging some parts of the code and making many helpful suggestions. R.B. wishes to
thank Rebecca Danos for useful discussions. R.B. is supported by an NSERC Discovery Grant and by funds from the Canada Research Chairs Program.

\end{document}